\documentclass[prd,preprint,tightenlines,floatfix,showpacs,preprintnumbers,nofootinbib,eqsecnum]{revtex4}

 \usepackage[dvips,final]{graphicx}
  \usepackage{amssymb}
   \usepackage{amsmath}
    \usepackage{amsfonts}
     \usepackage{epsfig}
      \usepackage{bm} 

\begin{document}


\title{On axial current in  rotating and accelerating medium}

\author{George Prokhorov$^1$}
\email{prokhorov@theor.jinr.ru}

\author{Oleg Teryaev$^{1,2,3}$}
\email{teryaev@theor.jinr.ru}

\author{V. I. Zakharov$^{2,4}$}
\email{vzakharov@itep.ru}
\affiliation{
{$^1$\sl 
Joint Institute for Nuclear Research, Dubna,  Russia \\
$^2$ \sl
Institute of Theoretical and Experimental Physics,
B. Cheremushkinskaya 25, Moscow, Russia \\
$^3$ \sl
National Research Nuclear University MEPhI (Moscow
Engineering Physics Institute), Kashirskoe Shosse 31, 115409 Moscow,
Russia \\
$^4$ \sl
School of Biomedicine, Far Eastern Federal University, 690950 Vladivostok, Russia
}\vspace{1.5cm}
}

\begin{abstract}
\vspace{0.5cm}
Statistical average  of the axial current is evaluated on the 
basis of the covariant Wigner function. In the resulting formula, 
chemical potential $\mu$, angular velocity  
$\Omega$  and acceleration $a$ enter in combination $\mu \pm (\Omega \pm ia)/2$.
 The limiting cases of zero mass and zero temperature are   investigated
in detail. 
In the zero-mass limit, the axial current is described by a smooth function only at 
temperatures higher than the Unruh temperature. At zero temperature, 
the axial current, as a function of the angular velocity and chemical potential, vanishes
 in a two-dimensional plane region.
\end{abstract}

\pacs{12.38.Mh, 11.30.Rd}

\maketitle

\section{Introduction}

 Recently many remarkable effects related to the properties 
of relativistic fluids have been discovered at theoretical level. 
The nature of these effects, on one hand, is associated with  
fundamental properties of matter, and, on the other hand, they 
are expected to be observable experimentally. Two best known examples of this kind 
are  the
Chiral Magnetic (CME) and the Chiral Vortical Effect (CVE)
manifested in electromagnetic and axial currents, respectively. 
For detailed discussion of the effects we refer the reader to the rich existing literature,
see in particular \cite{Vilenkin:1979ui, Vilenkin:1980zv, 
Vilenkin:1980fu, Kharzeev:2012ph, Fukushima:2008xe, 
Son:2009tf, Golkar:2012kb, Neiman:2010zi, Sadofyev:2010is, 
Zakharov:2012vv, Buzzegoli:2017cqy, Gao:2012ix, Gao:2017gfq, Landsteiner:2012kd}.

In \cite{Buzzegoli:2017cqy,Prokhorov:2017atp}  
the mean value of the axial current was calculated on the 
basis of the ansatz for the covariant Wigner 
function proposed in \cite{Becattini:2013fla}. The resulting formula reduces 
to the standard formula for the CVE in the approximation
linear in vorticity. 

Moreover, as is emphasized in \cite{Prokhorov:2017atp},
the entire series of expansion in the thermal vorticity can be 
summed up. The result
contains information on corrections to the standard
CVE. Some of these higher order terms  have been derived earlier within other approaches 
\cite{Vilenkin:1979ui, Vilenkin:1980zv}. Here
 we demonstrate that the  formula obtained can be greatly 
simplified and reduced to a form in which the angular velocity and acceleration enter as 
a real and imaginary chemical potentials,
respectively. Moreover, in
the zero-mass limit at temperatures below the Unruh temperature, 
additional  terms appear in the axial current, resulting in a
 jumplike behavior of the current, as a function of the temperature. 
According to \cite{Becattini:2017ljh} for linearly accelerated systems,
 the Unruh temperature is the lowest possible temperature. 
Our observation on existence of discontinuities in the behavior of the axial current
at temperatures below the Unruh temperature
 supports this conclusion. Note that existence of a boundary temperature 
proportional to the Unruh temperature 
was also derived in \cite{Florkowski:2018myy} starting
from the condition of positivity of energy density.

Evaluation of the axial current might have important 
phenomenological implications. Indeed,
the appearance of a significant  baryon polarization in 
heavy ion collisions can be one of most important 
experimental signatures of the CVE. In particular, papers in Refs.
\cite{Rogachevsky:2010ys, Sorin:2016smp, Baznat:2017jfj, Baznat:2017ars}, 
relate the polarization of baryons to an anomalous axial charge of quarks. 
On the other hand, the polarization effects can be investigated 
within the framework of the relativistic hydrodynamics of baryons 
\cite{Becattini:2017gcx, Becattini:2016gvu, Karpenko:2017lyj}, 
based on the Wigner function introduced in \cite{Becattini:2013fla}, 
from which the CVE can also be derived.  Note that the carriers of the axial charge 
differ in the two approaches. This situation served as a motivation
for us to study the effects in the axial current 
\cite{Becattini:2013fla, Prokhorov:2017atp}, connected with 
a finite mass of particles. An interesting phenomenon, 
which we find in this case, is the existence of a planar 
two-dimensional domain in the coordinates 
$\Omega$, $\mu$, where the axial current 
vanishes. Qualitatively, such a picture is 
associated with the above-mentioned observation that the 
angular velocity plays
the role of a new chemical potential.

The system of units $\hbar=c=k=1$ is used.

\section{Analysis of the effects of motion of the medium on the basis of the Wigner function}
\label{Sec:Wigner}

As is known, kinetic properties of a medium can be derived 
from the quantum field theory using the Wigner function, see, e.g., 
\cite{DeGroot:1980dk}. In the Ref. \cite{Becattini:2013fla} an ansatz for the Wigner function 
was proposed 
to describe  media with fermionic constituents in the state of a
local thermodynamic equilibrium.
Moreover, it was checked that the ansatz reproduces correctly  some known limiting cases. 
Based on this ansatz, 
the effects associated with thermal vorticity were investigated in various physical quantities
\cite{Becattini:2013fla, Buzzegoli:2017cqy, Prokhorov:2017atp,Florkowski:2018myy}. 
In particular, in \cite{Buzzegoli:2017cqy, Prokhorov:2017atp}, 
the axial current was first calculated, while an exact formula within the 
framework of this formalism was obtained in  \cite{Prokhorov:2017atp}. 

The Wigner function in \cite{Becattini:2013fla} is expressed in terms of the 
distribution function $X(x,p)$, which has the form of a modified Fermi-Dirac distribution
\begin{eqnarray}
&& X(x,p)=\Big(\exp[\beta_{\mu} p^{\mu}-\zeta]\exp\Big[-
\frac{1}{2}\varpi_{\mu\nu}\Sigma^{\mu\nu}\Big]+I\Big)^{-1}\,,
\label{X}
\end{eqnarray}
where $\zeta=\frac{\mu}{T}$, $\varpi_{\mu\nu}$ is the thermal vorticity tensor, 
and $\Sigma_{\mu\nu}=\frac{i}{4}[\gamma_{\mu},\gamma_{\nu}]$. 
Mean values of various physical quantities can be found by 
integrating the trace of the operator of the  quantity considered 
with the function $X(x,p)$ over the momentum space. 
Thus, for the axial current we have the following formula \cite{Becattini:2013fla}
\begin{eqnarray}
&&\langle j_{\mu}^{5}\rangle = 
-\frac{1}{16\pi^3} \epsilon_{\mu\alpha\nu\beta} \int \frac{d^3p}{\varepsilon} p^{\alpha}\Big\{
\mathrm{tr}\big(X\Sigma^{\nu\beta}\big) -
\mathrm{tr}\big(\bar{X}\Sigma^{\nu\beta}\big)\Big\}\,,
\label{axial mean}
\end{eqnarray}
where $\langle\cdot\rangle$ denotes  statistical averaging 
with normal ordering, $\bar{X}$ describes the contribution 
of  antiparticles and differs from (\ref{X}) in sign of $\zeta$ and $\varpi$. 
The matrix traces in (\ref{axial mean}) were exactly found in  \cite{Prokhorov:2017atp} in formula (4.3)
\begin{eqnarray}
&& \mathrm{tr}\big(X\Sigma^{\nu\beta}\big)=\Big\{\big(
\exp\big[(\beta p) -\zeta -\frac{g_{\omega}}{2T}+i \frac{g_a}{2T}\big]+1\big)^{-1} 
-\big(
\exp\big[(\beta p) -\zeta +\frac{g_{\omega}}{2T}-i \frac{g_a}{2T}\big]+1\big)^{-1}\Big\} \nonumber \\  
&& \frac{T}{2(g_{\omega}-i g_a)}[\varpi^{\nu\beta}
-i\,\mathrm{sgn}(\varpi_{\mu\alpha}\widetilde{\varpi}^{\mu\alpha})
\widetilde{\varpi}^{\nu\beta}]+ c.c. \,,
\label{trXSigma}
\end{eqnarray}
where $\widetilde{\varpi}^{\nu\beta}$ is the 
tensor dual to $\varpi^{\nu\beta}$, while 
$g_{\omega}$ and $g_a$ are scalar quantities that depend 
on acceleration $a^{\mu}=u^{\nu}\partial_{\nu}u^{\mu}$ 
and vorticity 
$\omega_{\mu}=\frac{1}{2}\epsilon_{\mu\nu\alpha\beta}u^{\nu}\partial^{\alpha}u^{\beta}$
\begin{eqnarray} 
&& g_{\omega}=\frac{1}{\sqrt{2}}\big(\sqrt{(a^2-\omega^2)^2+4(\omega a)^2}+a^2-
\omega^2 \big)^{1/2}\,, \nonumber \\
&& g_a=\frac{1}{\sqrt{2}}\big(\sqrt{(a^2-\omega^2)^2+4(\omega a)^2}-a^2+\omega^2 \big)^{1/2}\,.
\label{gwga}
\end{eqnarray}
Substituting (\ref{trXSigma}) into (\ref{axial mean}), we obtain
\begin{eqnarray}
&& \langle j_{\mu}^{5}\rangle =
\frac{\omega_{\mu}+i\, \mathrm{sgn}(\omega a) a_{\mu}}{2(g_{\omega}-i g_a)}
\int\frac{d^3 p}{(2\pi)^3}\Big\{
n_{F}(E_p-\mu - g_{\omega}/2 + i g_a/2)-\nonumber \\
&& n_{F}(E_p-\mu + g_{\omega}/2 - i g_a/2)  
+n_{F}(E_p+\mu - g_{\omega}/2 + i g_a/2)-\nonumber \\
&& n_{F}(E_p+\mu + g_{\omega}/2 - i g_a/2)
\Big\}+ c.c.\,,
\label{j5Wigner}
\end{eqnarray}
which is another form of Eq. (4.6) from \cite{Prokhorov:2017atp}.
Here $n_F(E)=(e^{E/T}+1)^{-1}$ is the Fermi-Dirac distribution,  
$a^{\mu}=u^{\nu}\partial_{\nu}u^{\mu}$ and 
$\omega_{\mu}=\frac{1}{2}\epsilon_{\mu\nu\alpha\beta}u^{\nu}\partial^{\alpha}u^{\beta}$ are
the four-acceleration and vorticity, respectively.

It is useful to consider a particular case by going into the comoving reference 
system and assuming that 
$\bold{\Omega}\,||\,\bold{a}$, that is, 
the acceleration 
directed along the rotation axis. 
Then $g_{\omega}=\Omega$, $g_a=a$, where $\Omega$ and $a$ 
are the moduli of three dimensional angular velocity and acceleration 
in the comoving frame, and (\ref{j5Wigner}) leads to
\begin{eqnarray}
&& \langle  \bold{j}^{5}\rangle =
\frac{1}{2} \int \frac{d^3 p}{(2\pi)^3}
\Big\{
n_{F}(E_p-\mu - \frac{\Omega}{2} + i \frac{a}{2})-
n_{F}(E_p-\mu + \frac{\Omega}{2} + i \frac{a}{2})+ \nonumber \\
&& n_{F}(E_p+\mu - \frac{\Omega}{2} + i \frac{a}{2})-
n_{F}(E_p+\mu + \frac{\Omega}{2} + i \frac{a}{2})+c.c.
\Big\}\,\bold{e}_{\,\Omega}\,,
\label{j5parallel}
\end{eqnarray}
where $\bold{e}_{\Omega}=\frac{\bold{\Omega}}{\Omega}$ 
is the unit vector in the direction of the angular velocity. 

Eq. (\ref{j5parallel}) demonstrates that
$\Omega$ and $a$ come  in a certain combination with the chemical potential.
Thus, the effect of rotation and acceleration reduces to a modification of the chemical
potential and introduction of a kind of an imaginary chemical potential.
This conclusion is worthy of further discussion.

In the limiting case of massless fermions, $m=0$, 
the integrals in (\ref{j5Wigner}) can be found analytically and 
expressed in terms of polylogarithms in the same way as was done 
in \cite{Prokhorov:2017atp}. Using the following property of the polylogarithms \cite{Prudnikov}
\begin{eqnarray} 
&& \mathrm{Li}_3(-e^{a+ib})-\mathrm{Li}_3(-e^{-a-ib})=-\frac{1}{6}\Big\{a+2\pi i\Big[\frac{b}{2\pi}-\lfloor\frac{b}{2\pi}+\frac{\mathrm{sgn}(b)}{2}\rfloor
\Big]\Big\}^3- \nonumber \\
&&\frac{\pi^2}{6}\Big\{a+2\pi i\Big[\frac{b}{2\pi}-\lfloor\frac{b}{2\pi}+\frac{\mathrm{sgn}(b)}{2}\rfloor
\Big]\Big\} \,,
\label{Li}
\end{eqnarray}
we obtain
\begin{eqnarray}
&& \langle j_{\mu}^{5}\rangle =\Big(\frac{1}{6}\big[T^2+\frac{a^{2} -\omega^2}{4\pi^2}\big]+\frac{\mu^2}{2\pi^2}\Big)\omega_{\mu}+\frac{1}{12\pi^2}(\omega a)\,a_{\mu}+\nonumber \\
&& \omega_{\mu}
\Big[
-\frac{4 \pi T g_a}{g_a^2+g_{\omega}^2}
(\frac{T^2}{6}+\frac{\mu^2}{2\pi^2}-\frac{g_a^2}{8\pi^2}-\frac{g_{\omega}^2}{8\pi^2})\lfloor\frac{g_a}{4\pi T}+\frac{1}{2}\rfloor -
2 T^2 \lfloor\frac{g_a}{4\pi T}+\frac{1}{2}\rfloor^2 +\nonumber \\
&&\frac{8\pi T^3 g_a}{3(g_a^2+g_{\omega}^2)}\lfloor\frac{g_a}{4\pi T}+\frac{1}{2}\rfloor^{3}
\Big]+
a_{\mu}\mathrm{sgn}(\omega a)
\Big[
-\frac{4 \pi T g_{\omega}}{g_a^2+g_{\omega}^2}
(\frac{T^2}{6}+\frac{\mu^2}{2\pi^2}+\frac{g_a^2}{8\pi^2}+\frac{g_{\omega}^2}{8\pi^2})\nonumber \\
&&\lfloor\frac{g_a}{4\pi T}+\frac{1}{2}\rfloor +
\frac{8\pi T^3 g_{\omega}}{3(g_a^2+g_{\omega}^2)}\lfloor\frac{g_a}{4\pi T}+\frac{1}{2}\rfloor^{3}
\Big]
\,,
\label{j5 full}
\end{eqnarray}
where $\lfloor\cdot\rfloor$ is the integer part. Note that in \cite{Prokhorov:2017atp} 
particular case $|\frac{b}{2\pi}+\frac{\mathrm{sgn}(b)}{2}|<1$ was considered 
under which formula (\ref{Li}) leads to the Eq. (4.9) from \cite{Prokhorov:2017atp}, 
which means that resulting formula Eq. (4.11) from \cite{Prokhorov:2017atp} corresponds 
to the case $T>\frac{g_a}{2\pi}$. Due to contributions from 
$\lfloor\frac{g_a}{4\pi T}+\frac{1}{2}\rfloor$ for $T<\widetilde{T}_U$, where $\widetilde{T}_U$ is
\begin{eqnarray}
&& \widetilde{T}_U=\frac{g_a}{2\pi}
\,,
\label{TU}
\end{eqnarray}
the formula (\ref{j5 full}) has discontinuities. For $T>\widetilde{T}_U$ 
the formula  (\ref{j5 full}) takes the form of Eq. (4.11) from \cite{Prokhorov:2017atp}, 
derived in approximation  $T>\widetilde{T}_U$
\begin{eqnarray}
\langle j_{\mu}^{5}\rangle =\Big(\frac{1}{6}\big[T^2+\frac{a^{2} -
\omega^2}{4\pi^2}\big]+\frac{\mu^2}{2\pi^2}\Big)\omega_{\mu}+\frac{1}{12\pi^2}(\omega a)\,a_{\mu}\,.
\label{j5Wigner m0}
\end{eqnarray}
It is interesting to note that in the case of $\bold{\Omega}\,||\,\bold{a}$ or $\bold{\Omega}=0$ 
the condition $T>\widetilde{T}_U$ results in $T>\frac{a}{2\pi}$, that is, 
the temperature is to  be greater than the Unruh temperature $T_{U}=\frac{a}{2\pi}$. 
The appearance of the Unruh temperature in Eq. (\ref{j5 full}) is a direct 
consequence of the fact that in (\ref{j5Wigner}) and (\ref{j5parallel}) the acceleration enters as an 
imaginary chemical potential. 
If both acceleration and angular velocity are nonzero and directed arbitrarily, 
the boundary temperature is generalized to $T_{U}\to\widetilde{T}_U(\Omega,a,\theta)$, 
where $\theta$ is the angle between $\bold{a}$ and $\bold{\Omega}$ in the comoving reference system.

According to \cite{Becattini:2017ljh}, the Unruh temperature 
is the minimum temperature that an accelerated medium can have. 
Apparently, this fact is the reason why the behavior of the axial 
current in (\ref{j5 full}) changes qualitatively at $T<\widetilde{T}_U$.
 A similar result on the existence of a boundary temperature 
proportional to the Unruh temperature on the basis of the same 
Wigner function \cite{Becattini:2013fla} was recently obtained 
in \cite{Florkowski:2018myy} by considering the energy-momentum 
tensor and the condition of positivity of the energy density. 
We note, however, that in \cite{Florkowski:2018myy} the boundary temperature is 
twice that of the Unruh temperature, 
which may be due to the fact that in \cite{Florkowski:2018myy} 
the Boltzmann limit was investigated \footnote{We are grateful to 
W. Florkowski, E. Speranza for pointing out the possibility of such an explanation.}.

Note that (\ref{j5Wigner m0}) in the first order in $\omega^{\mu}$ 
leads to the standard formula for CVE \cite{Buzzegoli:2017cqy, Prokhorov:2017atp}, 
while $(-\frac{\omega^2}{24\pi^2})\omega_{\mu}$ is consistent with the results of  
\cite{Vilenkin:1979ui, Vilenkin:1980zv} (see also \cite{Ambrus:2014uqa} 
for recent progress in the geometric approach, developed in \cite{Vilenkin:1979ui}).

\section{Effects of  finite mass}
\label{Sec:m}

There exist various approaches to calculating the 
polarization of baryons in heavy ion collisions. In particular, in the 
\cite{Rogachevsky:2010ys, Sorin:2016smp, Baznat:2017jfj, Baznat:2017ars} 
the axial charge of quarks, acquired by them due to the CVE, is considered, 
and this charge is associated with the polarization of baryons. 
On the other hand, in \cite{Becattini:2017gcx, Becattini:2016gvu, Karpenko:2017lyj}, 
the polarization is calculated on the basis of the Wigner function for a 
medium consisting of baryons, assuming equilibrium of the spin degrees of freedom.

Note that the CVE, which is essential for calculating the polarization 
in \cite{Rogachevsky:2010ys, Sorin:2016smp, Baznat:2017jfj, Baznat:2017ars}, 
arises in the approach of Refs. \cite{Becattini:2017gcx, Becattini:2016gvu, 
Karpenko:2017lyj} as well. However, in 
\cite{Rogachevsky:2010ys, Sorin:2016smp, Baznat:2017jfj, Baznat:2017ars}, 
quarks are considered as carriers of the axial charge, while in  
\cite{Becattini:2017gcx, Becattini:2016gvu, Karpenko:2017lyj} 
they are baryons, that is, particles with different masses. 
In view of this, it is useful to consider the effects of a finite mass in an axial current.

The most characteristic features in the behavior of 
the axial current arise at $T=0$. For simplicity, we also assume that $a_{\mu}=0$. 
Going into the comoving reference frame, we obtain $g_a=0$ and 
$g_W=\Omega$ in (\ref{gwga}). The integrals in (\ref{j5parallel}) can be evaluated 
analytically, and 
we get a simple formula
\begin{eqnarray}
&&\langle \bold{j}^{5}\rangle =\frac{1}{6\pi^2}\Big\{
\theta(\mu + \frac{\Omega}{2}-m)\big[(\mu+\frac{\Omega}{2})^2-m^2\big]^{3/2}- \nonumber \\
&&\theta(\mu - \frac{\Omega}{2}-m)\big[(\mu-\frac{\Omega}{2})^2-m^2\big]^{3/2}+ \nonumber \\
&&\theta(-\mu + \frac{\Omega}{2}-m)\big[(\mu-\frac{\Omega}{2})^2-m^2\big]^{3/2}- \nonumber \\
&&\theta(-\mu - \frac{\Omega}{2}-m)\big[(\mu+\frac{\Omega}{2})^2-m^2\big]^{3/2}
\Big\}\,\bold{e}_{\,\Omega}\,,
\label{j5T0}
\end{eqnarray}
where $\theta$ is the Heaviside function. From (\ref{j5T0}) it follows, 
in particular, that for $\Omega<2(m-|\mu|)$ the axial current is zero. 
This is in accord with the absence of chemical-potential effect if $\mu$ is 
smaller than the corresponding physical masses. Moreover, we find out that 
in case of a rotating medium, this is true for the
"effective" chemical potential incorporating the angular velocity.

\begin{figure*}[!h]
\centerline{\includegraphics[width=0.7\textwidth]{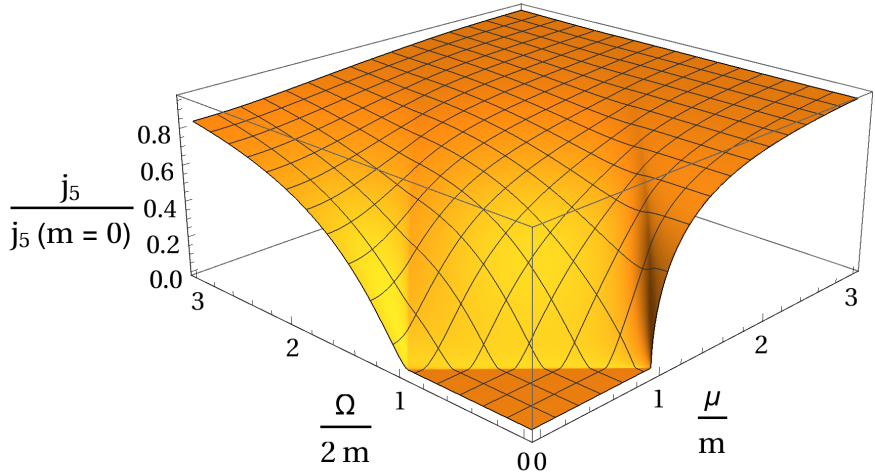}}\vspace{1cm}
\caption{Axial current (\ref{j5T0}), as a function of the chemical 
potential and angular velocity at zero temperature. The value of $|\bf{j}_5|$ 
 is normalized to its value (\ref{j5Wigner m0}) at zero mass.}
\label{fig:j5T0}
\end{figure*}

The behavior of  $j_5=|\bold{j_5}|$, see Eq. (\ref{j5T0}), as a function of 
$\Omega$ and $\mu$ is shown in Fig.\ref{fig:j5T0}. For 
$\Omega\gg m$ and $\mu\gg m$ the axial current asymptotically 
tends to its value at zero mass (\ref{j5Wigner m0}), $j_5(m=0)=
(\frac{\Omega^2}{24\pi^2}+\frac{\mu^2}{2\pi^2})\Omega$, 
as it should be. In general, due to the effects associated with the mass, 
$j_5$ in the 
massive case is always smaller than in the massless limit, as can be seen from Fig.\ref{fig:j5T0}.

\section{Discussion}
\label{Sec:Discussion}

Eq. (\ref{j5parallel}) exhibits some features which are challenging to explain theoretically.
Let us start with the ``imaginary acceleration'', $ia$.
Originally, see \cite{Becattini:2017ljh} and references therein, the acceleration $a$ enters 
the density operator $\hat{\rho}$ as a real number. Namely, in  absence of rotation:
\begin{equation}\label{added}
\hat{\rho}~=~(1/Z)\exp{\Big(-\hat{H}/T_0+a\hat{K}_z/T_0\Big)}~,
\end{equation}
where $\hat{H}$ is Hamiltonian and $\hat{K}_z$ is the generator of a Lorentz boost along the $z$ axis.
Note that $\ln\hat{\rho}$ is a Hermitian operator for real $a$.
In this sense the Eq. (\ref{added}) looks as a straightforward generalization
of the standard equilibrium density operator.

However, when applied to spinors in irreducible representations
the boost operators result in a complex number,
see, e.g.,  \cite{Ramond:1981pw}.
Indeed, the angular momentum $\hat{J}$ and boost generator $\hat{K}$ are combining to 
\begin{eqnarray}
\hat{N}=\hat{J}+i\hat{K},~~\hat{N}^{\dagger}~=~\hat{J}-i\hat{K}\,,
\end{eqnarray}
 where the eigenvalues  $N\neq 0, N^{\dagger} =0$
for left-handed spinors and $N^{\dagger}\neq 0, N=0$ for the right-handed spinors.

This leads to the different signs of acceleration of left and right fermions and could be called "chiral gravity". Axial current is a natural probe of such a solution.  

In this sense, the density matrix (\ref{added}) does not correspond to a 
genuine equilibrium if we stick to its interpretation in terms of flat space.
This is of no surprise, of course. Indeed, it is well known, for example,
that
in presence of an external gravitational field
the ordinary conservation of a current, $\partial_{\alpha}j^{\alpha}=0$, is becoming
a covariant conservation, $\nabla_{\alpha}J^{\alpha}=0$. Re-interpreted in terms of the flat
space the covariant conservation becomes a non-conservation, $\partial_{\alpha}j^{\alpha} \sim O(a)$. 
Similarly,
the expression for the divergence of the axial current obtained above gives
$\partial_{\alpha}j^{\alpha}\neq 0$ even in the limit of exact chiral symmetry if $a\neq 0$.

Note also that appearance of two signs of $ia$ may indicate the emergence of dissipating and 
unstable states which might also tunnel to each other. 

Turn now to the  ``modified chemical potential'', $\mu+\Omega/2$ emerging
in Eq. (\ref{j5parallel}). Note  that the possibility of considering 
the angular velocity as a chemical potential has already been 
noticed in the literature, see Ref. \cite{Vilenkin:1979ui}.
What we would like to emphasize here is that the coefficient $1/2$ in front of $\Omega$
can be interpreted as a consequence of the equivalence principle
according to which spin and angular momentum precess with
the same angular velocity
\cite{Teryaev:2016edw}.
In other words, the spin precession (for Dirac fermions) is twice slower than in the case of  magnetic field.
This factor of $1/2$, in turn, destroys the balance producing a zero mode in the
electromagnetic case. There is no zero mode in the gravitomagnetic field and, 
as a result, the axial anomaly in gravitational field is proportional 
to the curvature rather than connection. 

All these remarks can be considered as independent checks of  Eq. (\ref{j5parallel})
and support its validity.  

\section{Conclusions}
\label{Sec:Conclusions}

Basing on the ansatz for the Wigner function  proposed in \cite{Becattini:2013fla}, 
we obtained simple formulas for the axial current in the general case 
of massive fermions, see Eqs. (\ref{j5Wigner}) and (\ref{j5parallel}). In these formulas, 
the angular velocity and acceleration enter the Fermi-Dirac distribution in 
combination with the chemical potential. The zero-mass limit (\ref{j5 full}), 
(\ref{j5Wigner m0}), which is consistent in the linear approximation with 
the standard formula for the CVE, was studied. It is shown that in  case that
the acceleration and rotation are directed along the same axis, at a temperature 
lower than the Unruh temperature, the axial current has a series of discontinuities. 
In more general case of an arbitrary 
mutual orientation of the acceleration and angular velocity, 
the temperature (\ref{TU}) appears as a boundary, instead of the Unruh temperature.

Dependence of the axial current  on the mass of constituents
implied by Eq. (\ref{j5Wigner}) was investigated. 
In the limit $T=a_{\mu}=0$, (\ref{j5Wigner}) reduces to (\ref{j5T0}), and 
the axial current, as a function of the angular velocity and chemical potential, 
vanishes in the two-dimensional region $\Omega<2(m-|\mu|)$, as is shown in Fig.\ref{fig:j5T0}.

One can see, that the Wigner-function approach in the zero-mass limit reproduces, 
after the integration over momenta,  the anomaly induced contribution to the axial current, 
establishing the relation between different approaches to polarization.
One can even say, that the thermodynamical approach contains the "hidden anomaly".

{\bf Acknowledgments}

Useful discussions with F. Becattini, W. Florkowski, E. Grossi, A. Sorin, E. Speranza and A. Starobinsky  are gratefully acknowledged. 
The work was supported  by Russian Science Foundation
Grant No 16-12-10059.


\end{document}